# Amplification of Elliptically Polarized Sub-Femtosecond Pulses in IR-Field-Dressed Neon-Like Active Medium of a Plasma-Based X-ray Laser

I.R. Khairulin[1], V.A. Antonov[1,*], M.Yu. Ryabikin[1,2], M.A. Berrill[3],
V.N. Shlyaptsev[3], J.J. Rocca[3,4], and Olga Kocharovskaya[5]

[1]*Institute of Applied Physics of the Russian Academy of Sciences,
46 Ulyanov Street, Nizhny Novgorod 603950, Russia*
[2]*Lobachevsky State University of Nizhny Novgorod,
23 Prospekt Gagarina, Nizhny Novgorod 603950, Russia*
[3]*Department of Electrical and Computer Engineering, Colorado State University,
Fort Collins, CO80523, USA*
[4]*Department of Physics, Colorado State University,
Fort Collins, CO80523, USA*
[5]*Department of Physics and Astronomy, Texas A&M University,
College Station, 578 University Drive, Texas 77843-4242, USA*

We propose a method for amplifying a train of sub-femtosecond pulses of circularly or elliptically polarized extreme ultraviolet (XUV) radiation constituted by high-order harmonics of an infrared (IR) laser field, in a neon-like active medium of a plasma-based X-ray laser, additionally irradiated with a replica of a fundamental frequency IR field. It is shown that the ellipticity of the pulses can be maintained or increased during the amplification process. The experimental implementation is suggested in an active medium of an X-ray laser based on neon-like $Ti^{12+}$ ions irradiated by an IR laser field with a wavelength of 3.9 μm.

In recent years, considerable attention has been paid to the generation of elliptically and circularly polarized high-order harmonics (HHs) of optical radiation. The interest in this problem is due to the possibility to use an elliptically polarized radiation of extreme ultraviolet (XUV) and X-ray ranges for probing the magnetic [1,2] and chiral [3-5] media and processes in them [6,7], the development of spintronics [8], etc.

The straightforward approach to generating elliptically polarized HHs is to use an elliptically polarized laser field as a driver. Unfortunately, in this case the high-harmonic generation (HHG) efficiency decreases rapidly with increasing ellipticity of the fundamental laser field [9-11], and the ellipticity of harmonics steeply decreases with their order [9,12].

In order to overcome these limitations, several approaches were suggested, including (i) the resonant enhancement of elliptically polarized harmonics [13], (ii) conversion of linearly polarized harmonics to elliptically polarized ones via phase-shifting optics [14,15], (iii) generation of elliptically polarized harmonics in a gas of aligned molecules [16,17], as well as the use of (iv) cross-linearly-polarized two-color laser fields [18,19], and (v) bielliptic [20] or bicircular [21] fields, see [22] for the details. However, neither of them allows for the generation of sub-femtosecond XUV pulses with sufficiently high energy, high ellipticity and a well defined polarization state common for the harmonics of different orders. An amplification of 32.8 nm XUV radiation with circular polarization was experimentally shown in nickel-like $Kr^{8+}$ active medium of a plasma-based X-ray laser [23], but, in this case, only a single resonant harmonic was amplified.

Our recent work [24] has shown the possibility to amplify attosecond pulses formed by a set of linearly polarized HHs in a hydrogen-like active medium of a plasma-based X-ray laser dressed by a replica of a fundamental-frequency IR field with the same linear polarization. Under the action of an intense IR field, due to the linear Stark effect caused by the degeneracy of the $|2s\rangle$ and $|2p\rangle$ states in the hydrogen-like ions, the gain of the active medium is redistributed from the frequency of the inverted transition to a set of sidebands. However, because only two out of four energy degenerate states, corresponding to an excited energy level, experience linear Stark effect, in the hydrogen-like medium the gain redistribution occurs only for XUV/X-ray field with the same linear polariza-tion as that of the modulating field (we denote this direction by the z axis and propagation direction of both XUV and IR fields by the x axis), whereas for radiation with orthogonal polarization (along the y axis) the gain remains localized at the single resonance frequency, which prevents the amplification of elliptically or circularly polarized HHs.

In this work, we show that amplification of sub-femtosecond pulses of the HH radiation with an arbitrary elliptical and, in particular, circular polarization is possible in a neon-like active medium of a plasma-based X-ray laser modulated by a linearly polarized IR field. In this case, the IR field leads to a sub-laser-cycle harmonic modulation of energies of all the states of the resonant ions due to the quadratic Stark effect. As a result, the gain redistribution to combination frequencies occurs not only for XUV/X-rays with linear polarization, which coincides with the polarization of the IR field, but also for orthogonally polarized radiation, which makes it possible to amplify the set of circularly or elliptically polarized harmonics.

Below we consider the active medium of a plasma-based X-ray laser with inversion at the $3p^1S_0 \leftrightarrow 3s^1P_1$ transition of neon-like $Ti^{12+}$ ions with the unperturbed resonance wavelength of 32.6 nm [25-31]. The upper level is nondegenerate and corresponds to the state $|1\rangle$ with the total momentum $J=0$. The lower energy level is triply degenerate and corresponds to the $|2\rangle$, $|3\rangle$, and $|4\rangle$ states with $J=1$ and the momentum projection on the quantization axis $M=0$, $M=1$, and $M=-1$, respectively (see Fig.1 in [22]). Further, we assume that the active medium has the form of a thin cylinder oriented along the x axis, which is irradiated with an IR field polarized along the z axis:

$$\vec{E}_{IR}(x,t) = \vec{z}_0 E_M \cos\left[\Omega\left(t - x\sqrt{\varepsilon_{pl}^{(IR)}}/c\right)\right]. \quad (1)$$

Here $E_M$ and $\Omega$ are the amplitude and frequency of the IR field, $c$ is the speed of light in vacuum, $\varepsilon_{pl}^{(IR)} = 1 - \omega_{pl}^2/\Omega^2$ is the dielectric constant of the plasma for the IR field, $\omega_{pl} = \sqrt{4\pi N_e e^2/m_e}$ is the electron plasma frequency, $N_e$ is the concentration of free electrons in the plasma, $e$ and $m_e$ are the charge and mass of the electron, respectively. In Eq.(1), the pulse duration of the IR field is assumed to be significantly longer than the duration of all the processes under study, which allows us to consider it as a monochromatic one. Both the frequency of the IR field and its Rabi frequencies on the electric-dipole-allowed transitions from the states $|1\rangle$-$|4\rangle$ in $Ti^{12+}$ ions are much lower than the frequencies of these transitions. As a result, the major effect of the IR field on the $|1\rangle$-$|4\rangle$ states is a shift of the corresponding energy levels due to the quadratic Stark effect. In this case, the position of the i-th energy level (i=1,2,3,4) is determined by [32]

$$E_i(x,t) = E_i^{(0)} + \Delta_E^{(i)}\left(1 + \cos\left[2\Omega\left\{t - x\sqrt{\varepsilon_{pl}^{(IR)}}/c\right\}\right]\right), \quad (2)$$

where $E_i^{(0)}$ is the unperturbed energy value, $\Delta_E^{(i)} = \sum_{k \neq i}\left(\left|d_{ki}^{(z)}\right|E_M\right)^2/(2\hbar\omega_{ik})$ is the amplitude of the energy shift of the i-th state, $\omega_{ik}$ is the unperturbed frequency of the transition from the $|i\rangle$ state to the $|k\rangle$ state, $\hbar$ is Planck's constant, and $d_{ki}^{(z)}$ is the projection of the dipole moment of a given transition on the z axis. The summation in (2) is carried out over all the relevant states of the field-free ion (including the states $|1\rangle$-$|4\rangle$) [33]. We further introduce the notation $\Delta_\Omega^{(ij)} = \left(\Delta_E^{(i)} - \Delta_E^{(j)}\right)/\hbar$ for the amplitude of the frequency change of the $|i\rangle \leftrightarrow |j\rangle$ transition, where $i, j = 1,2,3,4$. For neon-like ions, both the average values and the amplitudes of the frequency shift of the $|1\rangle \leftrightarrow |2\rangle$ and $|1\rangle \leftrightarrow |3\rangle, |4\rangle$ transitions interacting with z- and y-polarized components of XUV radiation are different: $\Delta_\Omega^{(12)} \neq \Delta_\Omega^{(13)} = \Delta_\Omega^{(14)}$ as a result of different quadratic Stark shift of the states $|2\rangle$ and $|3\rangle, |4\rangle$.

In addition to the IR field, the active medium is irradiated with a set of its HHs of orders ranging from $2(q-k_{min})+1$ to $2(q+k_{max})+1$, which at the input to the medium, x=0, has the form

$$\vec{E}^{(inc)}(t) = \frac{1}{2}\sum_{k=k_{min}}^{k_{max}}\left(\vec{z}_0 \tilde{E}_{z,inc}^{(k)}(t) + \vec{y}_0 \tilde{E}_{y,inc}^{(k)}(t)\right)\exp\left\{-i(\omega + 2k\Omega)t\right\} + c.c., \quad (3)$$

where $\omega = (2q+1)\Omega$ is the carrier frequency of the HH field, $q$ is an integer, $\tilde{E}_{z,inc}^{(k)}(t)$ and $\tilde{E}_{y,inc}^{(k)}(t)$ are the slowly varying complex amplitudes of $z$- and $y$-polarization components of the field of the "$k$-th" harmonic with the frequency $\omega_k = \omega + 2k\Omega$, and c.c. denotes a complex conjugate.

The coupled space-time evolution of the HH field and the quantum state of Ti$^{12+}$ ions is described by the one-dimensional wave equation and the density-matrix equations, see [22]. In order to amplify both polarization components of the HH field, one should adjust the difference between the time-averaged frequencies of the transitions $|1\rangle\leftrightarrow|2\rangle$ and $|1\rangle\leftrightarrow|3\rangle,|4\rangle$ to an even multiple of the modulation frequency: $\bar{\omega}_{tr}^{(13)} - \bar{\omega}_{tr}^{(12)} = 2\Omega p$ (which is equivalent to $\Delta_E^{(3)} - \Delta_E^{(2)} = 2\hbar\Omega p$), where $p$ is an integer, by properly choosing the intensity of the IR field. In this case, if the carrier frequency of harmonics coincides with the time-averaged frequency of the transition $|1\rangle\leftrightarrow|2\rangle$, $\omega = \bar{\omega}_{tr}^{(12)}$, the analytical solution for the amplitudes of $z$- and $y$- polarization components of the "$k$-th" harmonic field takes the form [22]:

$$\tilde{E}_z^{(k)}(x,\tau) = E_{z,0}^{(k)}\theta(\tau)\exp\left\{g_k^{(z)}\left(P_\Omega^{(z)},\tau\right)x\right\}, \quad g_k^{(z)}\left(P_\Omega^{(z)},\tau\right) = g_{total}J_k^2\left(P_\Omega^{(z)}\right)\left(1-e^{-\gamma_{tr}\tau}\right), \quad (4a)$$

$$\tilde{E}_y^{(k)}(x,\tau) = E_{y,0}^{(k)}\theta(\tau)\exp\left\{g_{k-p}^{(y)}\left(P_\Omega^{(y)},\tau\right)x\right\}, \quad g_{k-p}^{(y)}\left(P_\Omega^{(y)},\tau\right) = g_{total}J_{k-p}^2\left(P_\Omega^{(y)}\right)\left(1-e^{-\gamma_{tr}\tau}\right), \quad (4b)$$

where $\tau = t - x\sqrt{\varepsilon_{pl}^{(XUV)}}/c$ is the local time, $\varepsilon_{pl}^{(XUV)} = 1 - \omega_{pl}^2/\omega^2$ is the dielectric constant of the plasma for the HH radiation, $\theta(\tau)$ is the Heaviside unit step function, $P_\Omega^{(z)} = \Delta_\Omega^{(12)}/(2\Omega)$ and $P_\Omega^{(y)} = \Delta_\Omega^{(13)}/(2\Omega)$ are the IR field-induced frequency modulation indices of the $|1\rangle\leftrightarrow|2\rangle$ and $|1\rangle\leftrightarrow|3\rangle,|4\rangle$ transitions, respectively, $g_k^{(z)}\left(P_\Omega^{(z)},\tau\right)$ and $g_{k-p}^{(y)}\left(P_\Omega^{(z)},\tau\right)$ are the effective gain coefficients for $z$- and $y$- polarization components of the HH field, $g_{total}$ is the gain factor in the absence of the Stark shift, $J_k(x)$ is the Bessel function of the first kind of order $k$, and $\gamma_{tr}$ is the decoherence rate at the transitions $|1\rangle\leftrightarrow|2\rangle$ and $|1\rangle\leftrightarrow|3\rangle,|4\rangle$. The analytical solution (4) implies that during the considered time interval, the population difference at the inverted transition is constant, the plasma is strongly dispersive for the modulating IR field so that there is no rescattering of HHs into each other, the spontaneous emission is negligible, and each harmonic in (3) is turned on at $t=0$ and then has a constant amplitude: $\tilde{E}_{z,inc}^{(k)}(t) = \theta(t)E_{z,0}^{(k)}$ and $\tilde{E}_{y,inc}^{(k)}(t) = \theta(t)E_{y,0}^{(k)}$, where $E_{z,0}^{(k)}$ and $E_{y,0}^{(k)}$ are complex numbers. In addition, the solution (4) implies an inertialess relationship between the resonant polarization of the medium and the XUV radiation (see [22,34]). The numerical results presented below were obtained without using any of the listed approximations.

As follows from Eqs.(4), the gain of the active medium is redistributed to multiple sidebands for both polarization components of the XUV field, and in the case $\bar{\omega}_{tr}^{(13)} - \bar{\omega}_{tr}^{(12)} = 2\Omega p$ the "$k$-th" induced gain line for $z$- polarization component of the XUV field overlaps with the gain line for $y$-polarization component numbered $k-p$. In the following, we assume $p=1$, as it is the easiest case for experimental implementation. In the Ti$^{12+}$ active medium the condition $\bar{\omega}_{tr}^{(13)} - \bar{\omega}_{tr}^{(12)} = 2\Omega$ corresponds to $P_\Omega^{(z)} \approx 12.57$ and $P_\Omega^{(y)} \approx 13.57$. The corresponding gain spectrum calculated via Eqs.(4) (see Fig.1) contains two different regions. On its left wing, particularly for $k=-9 - -13$ (red circles in Fig.1), the gain coefficients are nearly the same for $z$- and $y$- polarization components of the XUV field, which allows to amplify multiple harmonics preserving their polarization state. On the other hand, in the center of the spectrum, particularly for $k=-3, -1, 1, 3$, and $5$ (blue squares in Fig.1), $y$-polarization component of the XUV field is amplified stronger than $z$-polarization, which allows changing the polarization state of harmonics during their amplification.

In the following, we present the numerical results for the amplification of XUV pulse trains, formed by HHs of the modulating field, resonant to these two sets of the gain lines. We consider the IR field with wavelength 3.9 μm and intensity $8.26\times10^{16}$W/cm$^2$, which corresponds to

$\bar\omega_{tr}^{(13)} - \bar\omega_{tr}^{(12)} = 2\Omega$. The free electron and Ti$^{12+}$ ion densities in the active medium are $N_e$=5×10$^{19}$ cm$^{-3}$ and $N_{ion}$=4.2×10$^{18}$ cm$^{-3}$, respectively. The small signal gain in the absence of the IR field is 70 cm$^{-1}$. The coherence lifetime, $1/\gamma_{tr}$, nearly equals the collision time, $1/\gamma_{Coll} \approx 200$ fs. The radiative decay rates from the states |1⟩ and |2⟩-|4⟩ are $1/\Gamma_{rad}^{(1)} \approx 50.2$ ps and $1/\Gamma_{rad}^{(2,3,4)} \approx 3.34$ ps, respectively. Further, we plot the time dependencies of the intensities of the polarization components, $I_{z,y} = \frac{c}{8\pi}|\tilde E_{z,y}|^2$, where $\tilde E_y$ and $\tilde E_z$ are the slowly varying amplitudes of the total harmonic field, as well as the ellipticity, which in the considered case of π/2 phase shift between the polarization components is defined as $\sigma = |\tilde E_y|^2/|\tilde E_z|^2$, if $|\tilde E_y|^2 \le |\tilde E_z|^2$, and $\sigma = |\tilde E_z|^2/|\tilde E_y|^2$ otherwise.

Figure 2 shows the results for amplification of a train of circularly polarized pulses (the ellipticity σ=1 at *x*=0) formed by a combination of 129th, 131st, 133rd, 135th, and 137th harmonics of the modulating field (*k*=-13, -12, -11, -10, and -9, respectively). The pulse duration at the entrance to the medium is 1.2 fs, the pulse repetition period is 6.5 fs, the envelope of the pulse train has a FWHM duration 270 fs, and the central wavelength is 29.33 nm (the shorter pulses can be amplified with shorter wavelength modulating field, see [22]). After propagation through the active medium with the length *L*=1 cm the peak intensity and the total energy of the pulse train grow by 11.2 and 35.5 times, respectively. The pulse train is elongated (the bandwidth of each harmonic is reduced) due to amplification in optically dense medium, while the duration of each individual pulse is slightly increased due to nonuniform amplification of the harmonics of different orders. At the same time, the ellipticity of the XUV field is nearly preserved during the amplification process: σ=0.89 at the peaks of the most intense pulses from the amplified pulse train at τ ≈ 540 fs. The ellipticity variation within the modulation cycle is caused by the difference in ellipticity of the harmonics of different orders, while the overall reduction of ellipticity with increasing time (and increasing length of the medium) is due to different gain for *z*- and *y*-polarization components of the HH field.

Figure 3 illustrates the possibility to increase ellipticity of a train of sub-femtosecond pulses formed by a combination of the 149th, 153rd, 157th, 161st, and 165th harmonics of the modulating field (*k*=-3, -1, 1, 3, and 5 in Fig.1) during their amplification. In this case, the pulse duration at the entrance to the medium is 590 as, the pulse repetition period is 3.25 fs, the envelope has the same duration 270 fs, and the central wavelength is 24.85 nm. The field ellipticity at the entrance to the medium is σ(*x*=0,*τ*)=0.3. During the propagation through the medium, *y*-polarization component of the harmonic field grows faster than *z*-polarization, so that at *L*=1.1 cm their peak intensities are nearly equalized (while the phase difference between them remains unchanged). As a result, at the peak of the amplified pulse train (at τ ≈ 450 fs) the ellipticity reaches σ=0.995, which corresponds to almost circularly polarized radiation. At the same time, the ellipticity is nonuniform across the amplified pulse train: at the initial moments of time, *y*-polarization component is weaker than *z*-polarization, since some time is needed to establish the gain. At the peak of the amplified pulse train at τ≈450 fs, we have $|\tilde E_y|^2 \approx |\tilde E_z|^2$, while at its tail, *y*-polarization component dominates over *z*-polarization due to stretching in time caused by its stronger amplification. The increase in ellipticity of the pulse train is accompanied by increase in its total energy by 5.1 times. Similarly to Fig.2, each harmonic spectral line is narrowed (and the pulse train is stretched in time) due to high optical density of the medium, while the duration of sub-femtosecond pulses is slightly increased because of nonuniform amplification of the harmonics of different orders. The ellipticity of the amplified signal varies in time both on the scale of the XUV radiation envelope and within the IR field cycle. However, in the most energetic part of the pulse train the ellipticity is maximized at the peak of each sub-femtosecond pulse. Noteworthy, both in Figs.2 and 3 the amplification nearly preserves the pulse shapes.

In conclusion, in this paper we suggest an approach for the amplification of elliptically and circularly polarized sub-femtosecond XUV radiation pulses formed by HHs of an IR laser field, and increasing their ellipticity. It is proposed to use the active medium of neon-like Ti$^{12+}$ plasma-based

X-ray laser, simultaneously irradiated by the elliptically polarized HH field and the fundamental-frequency linearly polarized IR field. Due to the quadratic Stark shift caused by the IR field, the resonant energy levels of the ions oscillate in time and space with twice the IR field frequency, which results in redistribution of the gain of the active medium from the frequency of the inverted transition to the sidebands, separated from the resonance by even multiples of the modulation frequency. The gain redistribution occurs for both polarization components of the XUV radiation, parallel and perpendicular to the polarization of the IR field. With a proper choice of the intensity of the modulating field, the gain spectra for the orthogonal polarization components of the XUV radiation overlap, which makes it possible to amplify multifrequency elliptically polarized HH fields. The aforementioned matching of the gain spectra occurs, in particular, in a modulating field with a wavelength of 3.9 μm and an intensity of $8.26 \times 10^{16}$ W/cm$^2$. An important feature of the proposed method is its ability to amplify a set of harmonics while preserving their relative phases and approximately maintaining the temporal structure of the amplified signal. In particular, the possibility to amplify a train of circularly polarized pulses with a central wavelength of ~29 nm and an individual pulse duration of 1.2 fs is shown, while maintaining the polarization state and increasing the radiation energy by a factor of about 35. It is also shown that it is possible to amplify a train of pulses with a central wavelength of ~25 nm and a duration of 590 as with an increase in the ellipticity of radiation by more than 3 times (which corresponds to the transformation of elliptically polarized field into the circularly polarized one). In this case, the radiation energy increases by 5.1 times. The proposed method can be extended to other neon-like ions. In addition, due to the similarity of the energy structure of nickel-like and neon-like ions, this method can also be applied to the case of nickel-like active media (in particular, those based on $Mo^{14+}$, $Ag^{19+}$ or $Kr^{8+}$ ions). In prospect, this opens up the possibility of amplifying the radiation of harmonics of elliptical and circular polarization in shorter-wavelength spectral ranges [35].

The numerical calculations presented in this article were carried out with the support from the Russian Science Foundation (RSF, Grant No. 19-72-00140). The analytical studies were supported by the Center of Excellence «Center of Photonics» funded by The Ministry of Science and Higher Education of the Russian Federation, contract No. 075-15-2020-906. V.A.A. acknowledges personal support from the Foundation for the Advancement of Theoretical Physics and Mathematics BASIS. O.K. appreciates the support by the National Science Foundation (Grant No. PHY-2012194).

*Corresponding author: antonov@appl.sci-nnov.ru

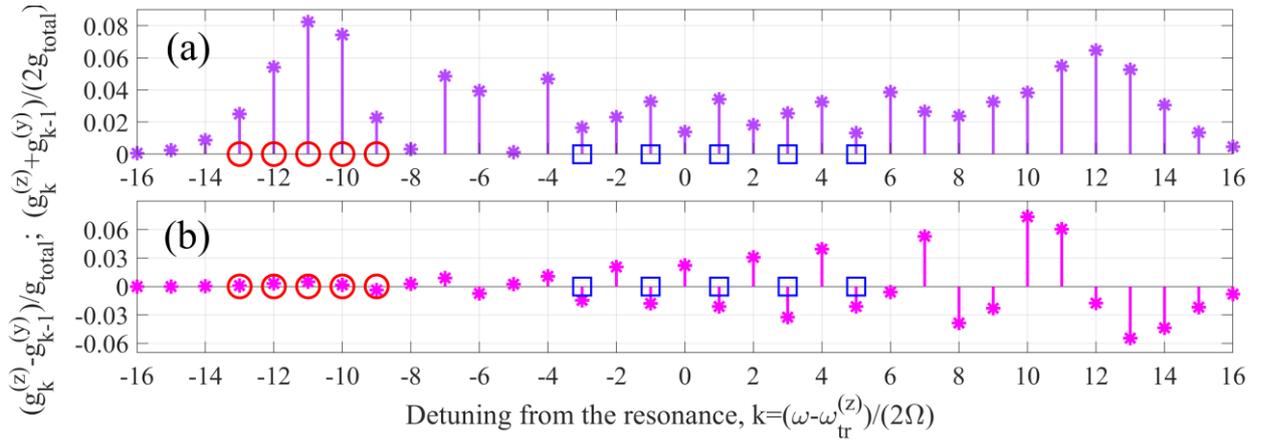

FIG. 1. (Color online) Half-sum (a) and difference (b) of the effective gain factors for the XUV radiation of $z$-polarization (4a) and $y$-polarization (4b). Red circles and blue squares mark the values of detuning, assumed in Figs.2 and 3, respectively.

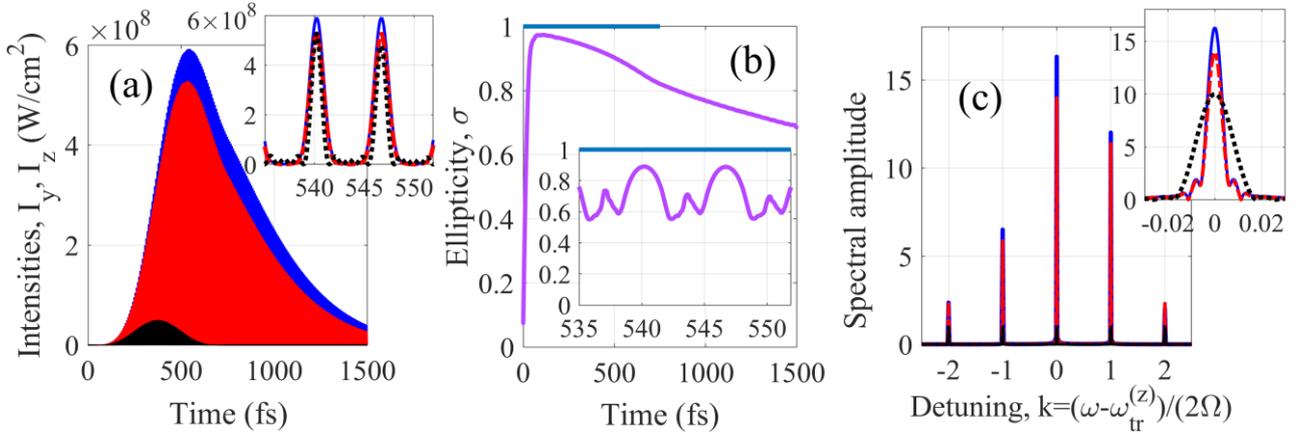

FIG. 2. (Color online) The result of amplification of the HHs resonant to the gain lines marked by red circles in Fig.1. (a) Time dependences of the intensities of the polarization components of the HH field. Black dotted curve and lowest intensity correspond to the incident field. Red dash-dot and blue solid curves are $y$- and $z$- components of the amplified field, respectively. In the inset, the intensity of the incident field is multiplied by factor 30 for visibility. (b) Ellipticity of the incident, blue line, and amplified, lavender curve, HH field at the maxima of individual pulses. Inset shows the ellipticity time dependence within the IR-field cycle. (c) Fourier transform of the polarization components of the incident field (black dotted curve) and amplified field (red dash-dot and blue solid curves showing $y$-and $z$-polarization components, respectively). Inset shows the shape of individual harmonic line (in the inset, the spectrum of the incident field is multiplied by factor 10 for visibility).

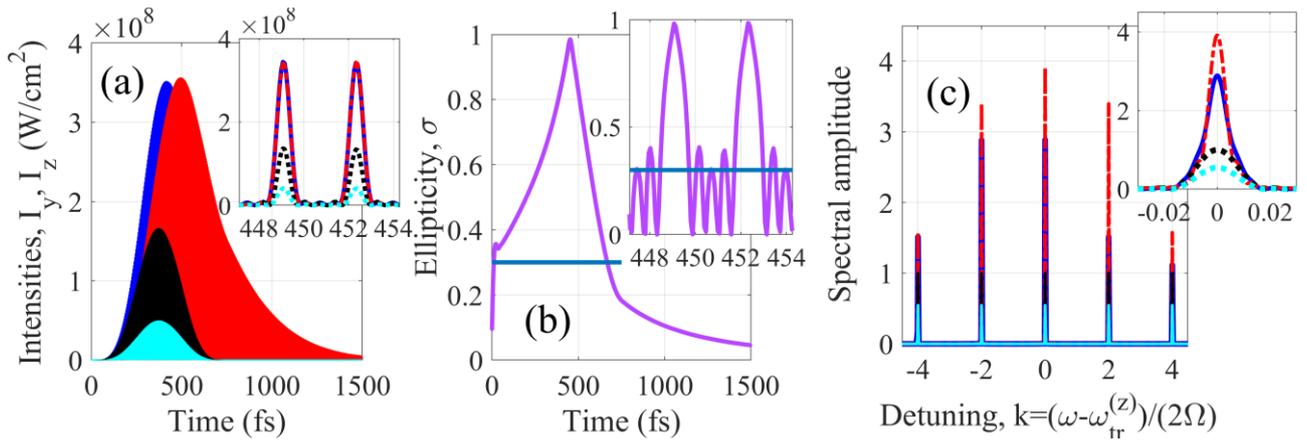

FIG. 3. (Color online) Same as in Fig.2 but for the HHs resonant to the gain lines marked by blue squares in Fig.1. Black and cyan dots correspond to *z*- and *y*- components of the incident field, respectively. In the insets in (a) and (c) the intensity and Fourier transform of the incident field are not zoomed.